# CONCEPT OF DYNAMIC MEMORY IN ECONOMICS

**Valentina V. Tarasova**,

Lomonosov Moscow State University Business School, Lomonosov Moscow State University, Moscow 119991, Russia; E-mail: v.v.tarasova@mail.ru;

**Vasily E. Tarasov**,

Skobeltsyn Institute of Nuclear Physics, Lomonosov Moscow State University, Moscow 119991, Russia; E-mail: tarasov@theory.sinp.msu.ru

**Abstract:** *In this paper we discuss a concept of dynamic memory and an application of fractional calculus to describe the dynamic memory. The concept of memory is considered from the standpoint of economic models in the framework of continuous time approach based on fractional calculus. We also describe some general restrictions that can be imposed on the structure and properties of dynamic memory. These restrictions include the following three principles: (a) the principle of fading memory; (b) the principle of memory homogeneity on time (the principle of non-aging memory); (c) the principle of memory reversibility (the principle of memory recovery). Examples of different memory functions are suggested by using the fractional calculus. To illustrate an application of the concept of dynamic memory in economics we consider a generalization of the Harrod-Domar model, where the power-law memory is taken into account.*

1. Introduction

Concept of memory is actively used not only in psychology, but also in the modern physics [1-13]. Economic processes with memory are actively studied in recent years (for example, see [14-24]). Fractional calculus and fractional differential equation [25-28], which are used derivatives and integrals of non-integer orders, are convenient tools to describe processes with memory in physical sciences (for example, see [11, 12] and references therein). In economics, the memory was first related to fractional differencing and integrating by Granger and Joyeux [20], and Hosking [21] in the framework of the discrete time approach [22, 23, 24]. Fractional differencing and integrating, which are suggested in these papers, are not directly connected with the fractional calculus or the well-known finite differences of non-integer orders [29]. In article [29] it was shown that the fractional differencing and integrating, which are proposed in [22, 23, 24], are the well-known Grunwald-Letnikov fractional differences. These differences have been suggested one hundred and fifty years ago. Recently fractional calculus has been used to describe economic processes with nonlocality in [30-35]. Fractional differential equations have been also applied for continuous-time finance in [36-56] by using the econophysics framework. These papers consider only the financial processes. The basic economic notions and the concepts of economic processes with memory are not considered. Economic processes with power-law memory have been considered in [57-75] in the framework of the continuous time approach. Using the fractional calculus as a mathematical tool to describe the power-law memory, we proposed generalizations of some basic economic concepts [57-75]. We have suggested the marginal value of non-integer order [59, 60, 61], the concepts of accelerator and multiplier with memory [62, 63], the elasticity [64] and the measures of risk aversion [65, 66] for processes with power-law memory, the methods of deterministic factor analysis [67]. The natural growth and logistic models, the Harrod-Domar and Keynes models have been generalized by taking account the power-law memory in [68-75]. These papers consider only simplest case of dynamic memory that is represented by the Riemann-Liouville fractional integrals and the Caputo fractional derivatives. This leads to questions about the applicability of other types of fractional derivatives and integrals to describe memory effects in economics. It is necessary to have a detailed consideration of the concept of dynamic memory for application in economics in the framework of the continuous time approach and fractional calculus.

Dynamic memory can be considered as an averaged characteristic (property), which describes the dependence of the process at a given time on the states in the past. Economic process with memory is a process, where economic parameters and factors (endogenous and exogenous variables) at a given time depend on their values at previous instants of a time interval. The economic process with dynamic memory assumes awareness of economic agents about the history of this process. In such processes, the behavior of economic agents is based not only on information on the state of the process $\{t, X(t)\}$ at a given moment of time t, but also on the use of information about the process states $\{\tau, X(\tau)\}$ at time instants $\tau \in [0, t]$. In economic processes a presence of memory means that there is an endogenous variable, which depends not only on values of an exogenous variable at present time, but also on its values at previous time points. A memory effect is related with the fact that the same change of the exogenous variable can leads to the different change of the corresponding endogenous variable. This leads us to the multivalued dependencies of these variables. The multivalued dependencies are caused by the fact that the economic agents remember previous changes of these variables, and therefore can react differently. As a result, an identical change of the exogenous variable may lead to the different dynamics of endogenous variables.

In this article we discuss a concept of dynamic memory and some mathematical methods, which allow us to describe economic processes with memory in the framework of the continuous time approach. The article consists of three parts. In the first part, we discuss the definition of the concept of dynamic memory. The second part deals with the general characteristics and principles of dynamic memory, which impose restrictions on the description of the memory. The third part provides examples of dynamic memory based on the use of fractional calculus. The last section of the article gives an example of the application of the concept of dynamic memory in macroeconomics.

## 2. Definition of Dynamic Memory

Economic models usually use two types of variables. At first, the exogenous variables (factors) are considered as independent (autonomous) variables, which are external to the model. Secondly, the endogenous variables (indicators) are internal variables of the model, which are formed inside the model. Indicators are described as variables that depend on the independent variables (factors). Thus endogenous variable can be considered as a response (reaction) to the action, which is described by the exogenous variable. In this paper, we will consider exogenous variables X(t) and endogenous variables Y(t) as real-valued functions of continuous time t.

The most general formulation of the economic process with memory is the following. In the economic process with memory there exists at least one endogenous variable Y(t) at the time t, which depends on the history of the change of X(τ) at $\tau \in (-\infty, t)$. Instead of the given verbal formulation we can use the following symbolic expression

$$Y(t) = F_{-\infty}^{t}(X(\tau)). \tag{1}$$

In equation (1) the symbol $F_{-\infty}^{t}$ denotes a certain method that allows us to find the value of Y(t) for any time t, if it is known X(τ) for $\tau \in (-\infty, t]$. We can say that $F_{-\infty}^{t}$ is an operator, which is a mapping from one space of functions to another. In continuum mechanics and physics $F_{-\infty}^{t}$ is called a functional, which transforms each history of changes of X(τ) for $\tau \in (-\infty, t]$ into the appropriate history of changes of Y(τ) with $\tau \in (-\infty, t]$. In mathematics, a functional is a map from a space of functions into its underlying field of scalars (numbers). For this reason we will use the mathematical term "operator".

In equation (1) the lower limit is taken as minus infinity. This means that the economic process is considered from the beginning of the creation of the Universe. Actually any economic process exists only during a finite time interval. Therefore, the lower limit can be select as start time of the process. Let us take this time as the reference point t=0 or $t = t_0$. As a result, expression (1) can be written in the form

$$Y(t) = F_0^t(X(\tau)). \tag{2}$$

The operator $F_0^t$ is said to be a linear operator, if the condition

$$F_0^t(a \cdot X_1(\tau) + b \cdot X_2(\tau)) = a \cdot F_0^t(X_1(\tau)) + b \cdot F_0^t(X_2(\tau)) \tag{3}$$

is satisfied for all a and b from the field of scalars.

In this paper, we consider only linear operators $F_0^t$ of a special kind. Namely, we will consider the Volterra operator, which is defined by the expression

$$F_0^t(X(\tau)) := \int_0^t M(t,\tau) \cdot X(\tau) \, d\tau. \tag{4}$$

The function M(t,τ), which is the kernel of the integral operator (4), is called the memory function. In this case, we can say that the dynamic memory is described by the Volterra operator $F_0^t$.

Expression (4) has a sense, if the integral (4) exists. Therefore the exogenous variable X(τ) is not necessarily a continuous function of time and the memory function can have an integrable singularity of some kind.

As a result, we can give the following definition of the concept of dynamic memory for the economic process.

**Definition of Dynamic Memory:** In economic process, dynamic memory is the property of the process, when there exists at least one endogenous variables Y(t), and an associated exogenous variable X(t), for which their dependence is described by the linear integral equation

$$Y(t) = \int_0^t M(t,\tau) \cdot X(\tau) d\tau, \tag{5}$$

where M(t,τ) is a function, which is called the memory function (or the linear response function).

It is obvious that not every equation of the form (5) can be used to describe the dynamic memory in the economic processes. Discussion of restrictions on equation (5) and consideration of the examples of memory function are the main aims of this paper.

Equation (5) can be interpreted as an equation of economic multiplier with dynamic memory, which is characterize by the function M(t, τ). We should note that the multiplier with power-law memory has been proposed in [62, 63].

Instead of the exogenous variable X(τ) of equation (5), we can consider the integer derivative $X^{(n)}(\tau) = d^n X(\tau)/d\tau^n$ of X(τ) with respect to time τ, where n is a nonnegative integer number. In this case, we have the equation

$$Y(t) = \int_0^t M(t,\tau) \cdot X^{(n)}(\tau) \, d\tau. \tag{6}$$

Equation (6) with n=1 can be interpreted as an equation of economic accelerator with dynamic memory, which is characterized by the function $M(t,\tau)$. Note that the accelerator with the power-law memory has been proposed in [62, 63].

The generalized linear accelerators with memory, which is characterized by the function $M(t,\tau)$, can also be defined by different integro-differential equations. For example, we can consider the equation

$$Y(t) = \frac{d^n}{dt^n} \int_0^t M(t,\tau) \cdot X(\tau)\, d\tau, \qquad (7)$$

and the equation

$$Y(t) = \frac{d^{n-k}}{dt^{n-k}} \int_0^t M(t,\tau) \cdot X^{(k)}(\tau)\, d\tau, \qquad (8)$$

where k=0, 1, …, n and n is a nonnegative integer number.

Keeping in mind a possibility of applying the concept of dynamic memory in economics, we first describe some general restrictions that can be imposed on the structure and properties of the memory function. These restrictions include: (1) the principle of fading memory; (2) the principle of memory homogeneity on time (the principle of non-aging memory); (3) the principle of memory reversibility (the principle of memory recovery). For simplicity we will not describe in detail the topological properties of the function spaces to which the functions X(t) and Y(t) belong, and the topological properties of operators $F_0^t$.

### 3. General Principles of Dynamic Memory

#### 3.1. Principle of fading memory

The first mathematical formulation of the principle of fading (dissipation) of memory has been proposed by Volterra [1]. A more complete definition of the principle of fading memory has been suggested in [2, 3, 4, 5]. The exact mathematical formulation of this principle is more complicated than that required for applications to economic theory in our paper. For the purposes of this article, which are bounded by the dynamic memory (5), we will use a somewhat simplified formulation of the principle of fading memory.

To describe the properties of the memory function and the fading memory, we consider an exogenous variable $X(\tau)$, which is different from zero on a finite time interval ($X(\tau) \neq 0$ for $\tau \in [0, T]$), and which is zero outside this interval ($X(\tau)=0$ for t>T). In this case, the factor $X(\tau)$ is represented through the Heaviside step function $H(T-\tau)$. Substituting $X(\tau) \cdot H(T-\tau)$ into equation (5) with times $t \in [T, \infty)$ instead of $X(\tau)$, we get the equation

$$Y(t) = \int_0^T M(t,\tau) \cdot X(\tau) d\tau. \qquad (9)$$

From equation (9) we see that for any given time t>T there is no impact (X(t)=0 for t>T), but the response is different from zero (Y(t)≠0 for t>T). This means that the memory about the impact, which existed during the time $\tau \in [0, T]$, is stored in the economic process. In other words, the economic process saves the history of changes of the exogenous variable. By the mean value theorem, there is a value $\xi \in [0, T]$ such that equation (9) can be represented in the form

$$Y(t) = M(t, \xi) \cdot X(\xi) \cdot T. \qquad (10)$$

We can consider an exogenous variable X(τ), which is expressed through the Dirac delta-function δ(τ–T). Substitution of X(τ)·δ(τ–T) into equation (5) instead of X(τ) gives the response Y(t) for t>T in the form $Y(t) = M(t, T) \cdot X(T)$. Using the notation $F_0^t(X(\tau))$, we can write $F_0^t(\delta(\tau - T)) = M(t, T)$ for the Volterra operators (4) with t>T.

As a result, we can see that the behavior of the endogenous variable Y(t), which is considered as a response on the impact (perturbation) X(τ), is determined by the behavior of the memory function M(t,ξ) or M(t,T), i.e. by the function M(t,τ) with fixed constant time τ. The behavior of the memory function M(t,τ) at infinite increase of t (t → ∞) and fixed τ determines the dynamics of the economic process with memory.

If the memory function tends to zero ($M(t, \tau) \to 0$) at $t \to \infty$, then the economic process completely forgotten the impact, which it has been subjected in the last time. In this case, the economic process, which is described by equation (5), is reversible (is repeated) in a sense. In other words, the memory effects did not lead to irreversible changes of the economic process and economic agents, since the memory about the impact has not been preserved forever. In the formulation of basic mathematical models of economic processes with memory, we can consider such memory functions that satisfy the following principle of fading memory.

**Principle of Fading Memory:** In economic process, dynamic memory, which is described by the Volterra operator (4), is fading if the memory function satisfies the condition $M(t, \tau) \to 0$ at $t \to \infty$ with fixed τ.

Using the symbolic representation of the dynamic memory, we can write the principle of fading memory in the form

$$\lim_{t \to \infty} F_0^t(\delta(\tau - T)) = 0. \qquad (11)$$

The dynamic memory will be called the memory with power-law fading if there is a parameter α>0 such that the limit $\lim_{t \to \infty} t^{-\alpha} \cdot M(t, \tau)$ is a finite constant for fixed τ.

If the memory function $M(t, \tau)$ tends to a finite limit at $t \to \infty$, the impact of exogenous variable X(t) on the economic process leads to the irreversible consequences in the sense that the memory of the impact is preserved forever.

Unbounded increase of the memory function $M(t,\tau)$ at $t \to \infty$ characterizes an unstable economic process with memory. This memory function cannot be used to describe stable economic processes. However, this type of functions can be applied in the models, which take into account the economic crises and economic shocks, when we can expect a manifestation of instability phenomena.

Let us consider equation (6) with $X(\tau)$, which is represented by the Dirac delta-function $\delta(\tau-T)$. Substitution of $X(\tau)=\delta(\tau-T)$ into equation (6) with integer $n \geq 1$ gives the response $Y(t)$ for $t>T$ in the form

$$Y(t) = (-1)^n \cdot \left(\frac{\partial^n M(t,\tau)}{\partial \tau^n}\right)_{\tau=T}. \tag{12}$$

Note that we can consider equation (5) with $X(\tau)$, which is represented by derivative of integer order $n \geq 1$ of the Dirac delta-function $\delta(\tau-T)$. Substitution of the variable $X(\tau) = \delta^{(n)}(\tau - T)$ into (5) gives (12).

Substitution of $X(\tau)=H(T-\tau)$ into equation (6) with integer $n \geq 1$ gives

$$Y(t) = (-1)^{n-1} \cdot \left(\frac{\partial^{n-1} M(t,\tau)}{\partial \tau^{n-1}}\right)_{\tau=T}. \tag{13}$$

Here we use that the first derivative of the Heaviside step function is the Dirac delta-function, where the derivative is considered in a generalized sense, i.e. on the space of test functions.

Note that in modern physics and mechanics the concept of fading memory assumes a set of stronger restrictions on memory function. For example, it is often assumed that the memory function tends to zero monotonically with increasing t. Let us explain this restriction in more detail. If the value of some endogenous variable $Y(t)$ is a linear function of the previous history of changes of an exogenous variable $X(t)$, the principle of fading memory states that the variable $Y(t)$ strongly depends on the changes in the recent history of $X(t)$, rather than on a distant history of the changes $X(t)$. In other words, the dependence of $Y(t)$ on the variable $X(\tau)$ at the previous times ($\tau<t$) is determined by using a memory function, which provides a continuously decreasing dependence on past events as they continued moving away from the considered points in time. In order to dependence of $Y(t)$ on the variable $X(\tau)$ satisfies the principle of fading memory it is sufficient to assume that the memory function $M(t,\tau)$ has been continuously decreasing function of time, i.e., $M(t_2, \xi) \leq M(t_1, \xi)$ for all $t_1$ and $t_2$ such that $t_2 > t_1 > 0$.

The principle of fading memory assumes that it is less probable to expect of strengthening of the memory with respect to the more distant events. However, in certain economic processes should take into account that the economic agents can remember sharp and significant changes of the exogenous variable $X(\tau)$, despite the fact that these changes were in the more distant past in comparison with the more weak changes in the nearest past. For this reason, it is acceptable to use the memory function, which is not monotonic decrease.

### 3.2. Principle of memory homogeneity on time

In physics the homogeneity of time means that equations of motion are invariant under the shift $t \to t + s$. Homogeneity of time means translational invariance with respect to time variable. Using the symbolic representation, we can write the principle of memory homogeneity on time in the form

$$F_{0+s}^{t+s}(X(\tau)) = F_0^t(X(\tau)). \tag{14}$$

Economic processes, for which this condition is satisfied for all variables t>0, can be interpreted as a process with the non-aging memory.

Homogeneity of time means that the flow of economic processes with memory in the same conditions, but at different times of their observation, occurs in the same way. If there are no designated points in time and the description of the economic process does not depend on the initial time point, then the memory function $M(t,\tau)$ satisfies the condition

$$M(t + s, \tau + s) = M(t, \tau). \tag{15}$$

Differentiating equation (15) with respect to the parameter s, and then assuming s=0, we obtain the equation

$$\partial M(t,\tau)/\partial t + \partial M(t,\tau)/\partial \tau = 0. \tag{16}$$

The general solution of this equation is an arbitrary function of t–τ, that is,

$$M(t,\tau) = M(t - \tau). \tag{17}$$

**Principle of Memory Homogeneity on Time:** In economic process, dynamic memory, which is described by the Volterra operator (4), is homogeneous on time if the memory function satisfies the condition $M(t,\tau) = M(t - \tau)$ for all τ and t≥τ .

In case of the Volterra operator (4), the memory that satisfies (14) or (17) can be interpreted as the non-aging memory. In the case (17), equation (5) is mathematically interpreted as the Duhamel convolution of functions, Y=M*X. Memory functions of the form (17) play an important role in the construction of models of economic processes, when the properties of processes do not depend on the choice of the time origin. For these processes, we can always choose the beginning of the reference time at t=0.

If the properties of economic process with memory do not depend on some selected points in time and, thus, does not depend on the choice of the reference time, the memory function will depend only on the difference t–τ. Here t–τ is the time that separates the present moment of time and the memorized event. Thus, the memory function for such processes can be represented in the form (17).

Let us consider equation (6) with $M(t,\tau) = M(t-\tau)$, where the factor $X(\tau)$ is represented by the Dirac delta-function $\delta(\tau–T)$ and the Heaviside step function $H(T–\tau)$. Substitution of $X(\tau)=\delta(\tau–T)$ into equation (6) gives the response $Y(t)$ for t>T in the form

$$Y(t) = \partial^n M(t,T)/\partial t^n, \tag{18}$$

where we use $\partial M(t,\tau)/\partial t = -\partial M(t,\tau)/\partial \tau$. Note that we can consider equation (5) with $X(\tau)$, which is represented by derivative of integer order n≥1 of the Dirac delta-function $\delta(\tau–T)$. Substitution the variable $X(\tau) = \delta^{(n)}(\tau - T)$ into (5) with $M(t,\tau) = M(t-\tau)$ gives (18). Substitution of $X(\tau)=H(T–\tau)$ into equation (6) gives $Y(t) = \partial^{n-1} M(t,T)/\partial t^{n-1}$. As a result, we can see that the behavior of the endogenous variable Y(t), which is considered as a response on the impact X(τ), is determined by the behavior of the n-th derivatives of the memory function $\partial^n M(t,T)/\partial t^n$ with fixed T.

There are economic processes with memory, which memory function cannot be represented in the form (17). For these economic processes the memory about the event depends not only on the time (t– τ), which separates the present moment of time t and the memorized event that occurred at time τ, but also on the state of the economic process in a moment of time τ. The initial point in time can be considered the time point of the process creation. If this time is considered as the reference point of time t = 0, then the time moment t> 0 can be interpreted as the age of process, and t = 0 can be interpreted as the time of the birth of process.

### 3.3. Principle of memory reversibility (memory recovery)

Until now we have assumed that the function X(t) is considered as an exogenous variable and we have interpreted it as the impact (or the perturbation). The function Y(t), which is an endogenous variable, is considered as a response (reaction) to the impact. In the general case, the equation $Y(t) = F_0^t(X(\tau))$ is irreversible, that is, does not exist an inverse operator $G_0^t$, for which $X(t) = G_0^t(Y(\tau))$, where t>0. In other words, if the function Y(t) is given on the interval (0,t), then it is not always possible to determine uniquely the function X(t) from equation $Y(t) = F_0^t(X(\tau))$.

**Principle of Memory Reversibility:** In economic process, dynamic memory, which is described by the equation $Y(t) = F_0^t(X(\tau))$ with the Volterra operator (4), is reversible if there is an operator $G_0^t$ such that $X(t) = G_0^t(Y(\tau))$ for all t>0.

In equation (5) we can consider Y(t) as a given function, and the variable X(t) can be considered as the unknown function. In this case, equation (5) is called the linear Volterra equations of the first kind. Existence of solutions of this equation, i.e. the reversibility of equation (5), has been studied in the theory of integral equations.

It should be noted that the question of the reversibility of equations (5) and (6) are connected with the principle of duality of accelerator with memory and multiplier with memory, which is proposed in [62] for the power-law memory.

Equations (5) and (6) can be associated with the so-called inverse problems of dynamics, which is the economic dynamics in our case. For example, if we know the memory function $M(t,\tau)$ of an economic process and an endogenous variable $Y(t)$, then we can consider the problem of finding (restoration) of the impact function $X(t)$, which is the exogenous variables of this process. This type of problems includes a finding of solutions for the equations of macroeconomic growth models with memory. For power-law memory such problems have been solved for the natural growth model [68, 69, 70, 71], the hereditary Harrold-Domar model [72, 73] and the hereditary Keynes model [74, 75].

Another inverse problem is related to econometrics ("hereditary econometrics"). The known functions of the exogenous variable $X(t)$ and the endogenous variable $Y(t)$ can be used to determine the memory function $M(t,\tau)$ of the considered economic process. Note that this problem has a great importance for the study of properties of dynamic memory in the real economic processes. In the framework of the model with the one-parameter power-law memory, this problem has been considered in [76] in the form of determination of fading order $\alpha$.

## 4. Examples of Dynamic Memory

### 4.1. Absence of memory (amnesia) and infinitesimal memory

For processes without dynamic memory, the memory functions are expressed in terms of the Dirac delta-function in the form

$$M(t,\tau) = M(t-\tau) = m(\tau) \cdot \delta(t-\tau), \tag{19}$$

where $\delta(t-\tau)$ is the Dirac delta-function. The absence of memory means that the endogenous variable $Y(t)$, which is the economic indicator, is determined by an exogenous variable $X(t)$, which is the factor, only at the moment of time t. We can say that instantaneous forgetting of the history of the factor changes is realized. Substitution of (19) into equation (5) gives

$$Y(t) = m(t) \cdot X(t). \tag{20}$$

Equation (20) is the equation of economic multiplier, which describes process without dynamic memory and time delay (time lag). This process connects the sequence of subsequent states of the economic process to the previous state only through the current state for each time t.

Instead of the exogenous variable $X(\tau)$ we can consider the integer derivative $X^{(n)}(\tau) = d^n X(\tau)/d\tau^n$ of the factor $X(\tau)$ with respect to time $\tau$, where n is an positive integer number. In this case we can use equation (6). Then equation (6) with memory function (19) gives the equation

$$Y(t) = m(t) \cdot X^{(n)}(t). \qquad (21)$$

Equations (21) with n=1 is the equation of the standard accelerator without memory.

It should be noted that we cannot consider the multiplication of the Dirac delta-functions since these functions are the generalized functions, which are treated as functionals on a space of test functions. Therefore we cannot substitute expression (19) and the factor $X(\tau)=\delta(\tau–T)$ into equations (5) and (6). The factor $X(\tau)=\delta(\tau–T)$ can be substituted into equation (20) and (21) only.

It is known that the derivatives $X^{(n)}(t)$ of the integer order n are determined by the values of $X(\tau)$ in the infinitesimal neighborhood of the time point $\tau=t$. In case (21), we can use the term of infinitesimal memory, when the forgetting occurs in an infinitely small time interval. In fact, the equations with derivatives of integer orders describe a process in which all economic agents have infinitely fast amnesia. In other words, the economic models, which use only integer derivatives, can be useful, when economic agents forget the history of changes of the economic factors during an infinitely small integral of time. As a result, the differential equations with derivatives of integer orders with respect to time cannot describe the economic processes, for which the memory effect is manifested on a finite time interval.

If the memory function $M(t,\tau) = M(t–\tau)$ has the form $M(t)=m\cdot\delta(t–T)$, then equation (3) can be written as an equation of multiplier with is a fixed-time delay of T periods, where the time-constant of the lag T is a given positive integer: $Y(t) = m \cdot X(t − T)$, [77, p. 23].

### 4.2. Complete (perfect, ideal) memory

Let us consider a finite time interval [0,t], in which the factor X(t) influences on the indicator Y(t). Using the symbolic representation, we can consider the condition on dynamic memory in the form

$$F_0^t(1) = 1. \qquad (22)$$

Economic processes, for which condition (22) is satisfied for all t>0, can be interpreted as a unit preserving memory. For the linear operators $F_0^t$, this condition means that constant exogenous variable does not change. In other words, the memory remembers unchanged variable infinitely long. For the Volterra operator, condition (22) leads to the following restriction on the memory function

$$\int_0^t M(t,\tau)d\tau = 1. \qquad (23)$$

Condition (23) means that equation (5) should give Y(t)=1 for X(t)=1.

If the memory is non-aging memory, i.e. $M(t,\tau)=M(t–\tau)$, then condition (23) has the form

$$\int_0^t M(\tau)d\tau = 1. \qquad (24)$$

The function M(t,τ)= M(t–τ), which satisfies the normalization condition (24) is often called the weighting function [77, p. 26]. This type of memory functions M(t) is often used to describe economic processes with continuously distributed lag [77, p. 25]. The existence of the time delay (lag) is connected with the fact that the processes take place with a finite speed, and the change of the economic factor does not lead to instant changes of indicator that depends on it.

If condition (22) or (23) holds, the economic process goes through all the states without any losses. In this case we say that the memory function describes the complete (perfect, ideal) memory.

### 4.3. Memory with power-law fading

One of the first who considered the natural processes with fading memory was Vito Volterra. He proposed the principle of fading memory [1, p. 227]. Volterra used the integral equations to describe the processes with memory [1, p. 226-229], which he called hereditary processes. In modern physics, the fading memory [2, 3, 4, 5, 6] plays an important role in the description of various physical processes, [7, 8, 9, 10, 11, 12, 13].

To describe the economic processes with fading memory, we can use the equations with derivatives and integrals of non-integer orders [25], which are a special case of integro-differential equations, [26, 27, 28]. To describe the dynamic memory with power-law fading, we can use the memory function in the form

$$M(t, \tau) = M_\alpha(t - \tau) = \frac{1}{\Gamma(\alpha)} \frac{m}{(t-\tau)^{1-\alpha}}, \tag{25}$$

where $\Gamma(\alpha)$ is the Gamma function, α>0 is a parameter that characterize the power-law fading, t>τ, and m is a positive real number, which can depend on α in general. Substitution of expression (25) into equation (5) gives the fractional integral equation of the order α>0 in the form

$$Y(t) = m \cdot (I_{0+}^\alpha X)(t), \tag{26}$$

where $I_{0+}^\alpha$ is the left-sided Riemann-Liouville integral of the order α>0 with respect to time variable. This integral is defined [27, p. 69-70] by the equation

$$(I_{0+}^\alpha X)(t) := \frac{1}{\Gamma(\alpha)} \int_0^t \frac{X(\tau)d\tau}{(t-\tau)^{1-\alpha}}, \tag{27}$$

where $\Gamma(\alpha)$ is the Gamma function and t>0. In equation (27) the function X(t) is assumed to be measurable on the interval (0,T) and it must satisfy the condition $\int_0^t |X(\tau)|d\tau < \infty$. The Riemann-Liouville integral (27) is a generalization of the standard integration [25]. Note that the Riemann-Liouville integration (27) of the order α = 1 gives the standard integration of first order, $(I_{0+}^1 X)(t) := \int_0^t X(\tau)d\tau$.

Equation (26) describes a multiplier with power-law memory [62], and the parameter m is the coefficient of the multiplier.

If we consider a unit exogenous variable, i.e. X(τ)=1 for τ ∈ [0, ∞), then equation (26) gives the endogenous variable in the form

$$Y(t) = \frac{m}{\Gamma(\alpha)} \cdot \int_0^t (t-\tau)^{\alpha-1} d\tau = \frac{m \cdot t^\alpha}{\Gamma(\alpha+1)}. \tag{28}$$

Therefore, the memory, which is defined by memory function (25), cannot be considered as a unit preserving memory. This mean that processes cannot remember a constant value during the infinitely long period of time.

Let consider an exogenous variable X(τ), which is represented by the Heaviside step function H(T–τ), i.e. X(τ) = 1 for τ ∈ [0, T] and X(τ)=0 for τ ∈ [T, ∞). Substitution of X(τ)=H(T–τ) into equation (27) with t ∈ [T, ∞), we get

$$Y(t) = \frac{m}{\Gamma(\alpha+1)}(t^\alpha - (t-T)^\alpha), \tag{29}$$

where we use $\alpha \cdot \Gamma(\alpha) = \Gamma(\alpha+1)$. We see that for t>T, when there is no impact (X(t)=0 for t>T), the response is different from zero (Y(t)≠0 for t>T) and it has the power-law form. If we consider an exogenous variable X(τ), which is represented by the Dirac delta-function δ(τ–T), then equation (26) gives the response for t>T in the form

$$Y(t) = \frac{m}{\Gamma(\alpha)} \cdot (t-T)^{\alpha-1}. \tag{30}$$

As a result, the behavior of the response Y(t) on the impact X(τ) has the power-law type. Therefore the memory, which is defined by memory function (25), can be considered as fading memory, for which the fading has the power-law form.

If we use the exogenous variable

$$X(\tau) = x(\tau) \cdot \sum_{k=1}^\infty \delta\left(\frac{\tau}{T} - k\right), \tag{31}$$

where x(τ) is continuous at all points t=kT, then the response for t: N·T <t<(N+1)·T takes the form

$$Y(t) = \frac{m}{\Gamma(\alpha)} \cdot \sum_{k=0}^N (t - k \cdot T)^{\alpha-1} \cdot x(k \cdot T). \tag{32}$$

Using the notations

$$Y_N := Y(N \cdot T - 0) = \lim_{\varepsilon \to 0+} Y(N \cdot T - \varepsilon), \tag{33}$$

$$x_k := x(k \cdot T - 0) = \lim_{\varepsilon \to 0+} x(k \cdot T - \varepsilon), \tag{34}$$

we get the discrete map with memory

$$Y_{N+1} = \frac{m \cdot T^{\alpha-1}}{\Gamma(\alpha)} \cdot \sum_{k=0}^N (N + 1 - k)^{\alpha-1} \cdot x_k. \tag{35}$$

Using equation (35) with replacement N+1 by N, and subtraction the result from equation (35), we get the discrete map with memory in the form

$$Y_{N+1} = Y_N + \frac{m \cdot T^{\alpha-1}}{\Gamma(\alpha)} \cdot x_N + \frac{m \cdot T^{\alpha-1}}{\Gamma(\alpha)} \cdot \sum_{k=0}^{N-1} V_\alpha(N-k) \cdot x_k, \tag{36}$$

where $V_\alpha(z)$ is defined by $V_\alpha(z) := (z+1)^{\alpha-1} - (z)^{\alpha-1}$.

This equation describes the discrete multiplier with power-law memory. The discrete economic accelerator with power-law memory has been suggested in [63, 71].

Equation (26) with α=1 takes the form $Y(t) = m \cdot \int_0^t X(\tau)d\tau$. The action of the derivative of first order on this equation gives the standard equation of the accelerator $X(t) = a \cdot Y^{(1)}(t)$, where $a = 1/m$ is the accelerator coefficient.

Let us consider the integer derivative $X^{(n)}(\tau) = d^n X(\tau)/d\tau^n$ of the factor $X(t)$ as an exogenous variable of equation (26). In this case, we will use the memory function (25) is the form

$$M(t,\tau) = M_{n-\alpha}(t-\tau) = \frac{1}{\Gamma(n-\alpha)} \frac{a}{(t-\tau)^{\alpha-n+1}}, \tag{37}$$

where n:=[α]+1. Then we get the equation

$$Y(t) = a \cdot (D_{0+}^{\alpha} X)(t), \tag{38}$$

where $D_{0+}^{\alpha}$ is the left-sided Caputo fractional derivative of the order α≥0 with respect to variable t [27, p. 92]. This Caputo fractional derivative is defined by the equation

$$(D_{0+}^{\alpha} X)(t) := \frac{1}{\Gamma(n-\alpha)} \int_0^t \frac{X^{(n)}(\tau)d\tau}{(t-\tau)^{\alpha-n+1}}, \tag{39}$$

where $\Gamma(\alpha)$ is the Gamma function, t>0, and $X^{(n)}(\tau)$ is the derivative of the integer order n:=[α]+1 with respect to τ. It is assumed that the function $X(\tau)$ has derivatives up to the (n-1)th order, which are absolutely continuous functions on the interval [0,t].

Equation (38) describes the equation of economic accelerator with memory with the power-law fading of the order α≥0, where a is the acceleration coefficient. The concept of the accelerator with memory has been proposed in [62].

Let us consider equation (38), where the factor $X(\tau)$ is represented by the Dirac delta-function δ(τ–T) and the Heaviside step function H(T–τ). Substitution of $X(\tau)=\delta(\tau-T)$ into equation (38) gives the response Y(t) for t>T in the form

$$Y(t) = \frac{\partial^n M_{n-\alpha}(t-T)}{\partial t^n} = \frac{a}{\Gamma(-\alpha)} \cdot (t-T)^{-\alpha-1} \tag{40}$$

for non-integer values of order α. Substitution of $X(\tau)=H(T-\tau)$ into equation (38) gives

$$Y(t) = \frac{\partial^{n-1} M_{n-\alpha}(t-T)}{\partial t^{n-1}} = \frac{a}{\Gamma(-\alpha)} \cdot (t-T)^{-\alpha}, \tag{41}$$

where α is non-integer. As a result, we can see that the behavior of the endogenous variable Y(t), which is considered as a response on the impact $X(\tau)$, has the power-law type.

**4.4. Memory with multi-parametric power-law fading.**

Note that we have considered the simplest models of dynamic memory, in which the power-law fading is characterized by one parameter α only. In economic models we can take into account the presence of different types of memory fading, which characterize the different types of economic agents. In this case, to describe fading memory we can use the memory function M(t,τ) as a sum of functions (25) with different parameters of fading memory. For example, we can use the following two-parameter description of the fading memory

$$M(t,\tau) = M_{\alpha,\beta}(t-\tau) = \frac{1}{\Gamma(1-\alpha)}\frac{m_\alpha}{(t-\tau)^\alpha} + \frac{1}{\Gamma(1-\beta)}\frac{m_\beta}{(t-\tau)^\beta}, \quad (42)$$

where α>0, β>0, t>τ, and $m_\alpha$, $m_\beta$ are numerical coefficients. Substitution of expression (42) into equation (5) gives

$$Y(t) = m_\alpha \cdot (I_{0+}^\alpha X)(t) + m_\beta \cdot \left(I_{0+}^\beta X\right)(t). \quad (43)$$

In equation (43), we can use the integer derivatives $X^{(n)}(\tau) = d^n X(\tau)/d\tau^n$ and $X^{(m)}(\tau) = d^m X(\tau)/d\tau^m$, where n=[α]+1 and m=[β]+1 are positive integer numbers, as exogenous variables. In this case, the memory function (42) can be considered in the form

$$M(t-\tau) = M_{n-\alpha,m-\beta}(t-\tau) = \frac{1}{\Gamma(n-\alpha)}\frac{a_\alpha}{(t-\tau)^{\alpha-n+1}} + \frac{1}{\Gamma(m-\beta)}\frac{a_\beta}{(t-\tau)^{\beta-m+1}}. \quad (44)$$

As a result, we get the equation

$$Y(t) = a_\alpha \cdot (D_{0+}^\alpha X)(t) + a_\beta \cdot \left(D_{0+}^\beta X\right)(t). \quad (45)$$

In order to take into account N different parameters of memory fading, we can use the equation

$$Y(t) = \sum_{k=1}^N a_k \cdot \left(D_{0+}^{\alpha_k} X\right)(t). \quad (46)$$

Using the memory with multi-parametric power-law fading, we can consider economic models that allow us to describe the memory effects in economics with different type of economic agents.

### 4.5. Memory with power-law fading of variable order

In some economic processes with memory, the parameter α of memory fading can be changed during the time, i.e. α=α(t). In this case, we can consider the memory function (25) in the form

$$M(t,\tau) = M_{\alpha(t)}(t-\tau) = \frac{1}{\Gamma(\alpha(t))}\frac{m}{(t-\tau)^{1-\alpha(t)}}, \quad (47)$$

where the variable order α(t)≥0. Substituting (47) into equation (5) we obtain the integral equation in the form

$$Y(t) = m \cdot (I_{0+}^{\alpha(t)} X)(t), \quad (48)$$

where $I_{0+}^{\alpha(t)}$ is the Riemann-Liouville fractional integral of variable order [78, 79, 80, 55]. If α(t) = α, then equation (48) gives (26).

Using $X^{(n)}(\tau) = d^n X(\tau)/d\tau^n$ with n=[α]+1 instead of X(τ), and the memory function $M(t,\tau) = M_{n-\alpha(t)}(t-\tau)$ instead of $M(t,\tau) = M_{\alpha(t)}(t-\tau)$, we get the equation of accelerator of variable order

$$Y(t) = a \cdot (D_{0+}^{\alpha(t)} X)(t), \quad (49)$$

where $D_{0+}^{\alpha(t)}$ is the left-sided Caputo fractional derivative of the variable order α(t)≥0. For α(t) = α equation (49) gives (38).

**4.6. Memory with generalized power-law fading**

Dynamic memory can be characterized by more complex fading behavior than those that we have considered in sections III.1-III.5. In this case the dynamic memory should be described by more complex memory function. In order to describe the economic dynamics with more complex memory, we can use the generalized fractional calculus [81, 82]. Let us give some examples of the generalized memory functions M(t,τ).

For example, we can consider the memory function in the form

$$M(t,\tau) = \frac{t^{-\alpha-\eta}}{\Gamma(\alpha)} \frac{m \cdot \tau^{\eta}}{(t-\tau)^{1-\alpha}}, \qquad (50)$$

where η is the real number and α>0. Substituting (50) into equation (5), we obtain the integral equation of the order α>0 in the form

$$Y(t) = m \cdot (I_{0+;\eta}^{\alpha} X)(t), \qquad (51)$$

where $I_{0+;\eta}^{\alpha}$ is the left-sided Kober fractional integral of the order α>0 with respect to time variable [27, p. 106].

We can consider the memory function in the form

$$M(t,\tau) = M_{\sigma}^{\alpha,\eta}(t,\tau) := \frac{\sigma \cdot t^{-\sigma(\alpha+\eta)}}{\Gamma(\alpha)} \frac{m \cdot \tau^{\sigma(\eta+1)-1}}{(t^{\sigma}-\tau^{\sigma})^{1-\alpha}}, \qquad (52)$$

where σ>0, η is the real number and α>0. For σ=1, expression (52) takes the form (50). Substituting (52) into equation (5), we obtain the integral equation in the form

$$Y(t) = m \cdot (I_{0+;\sigma,\eta}^{\alpha} X)(t), \qquad (53)$$

where $I_{0+;\sigma,\eta}^{\alpha}$ is the left-sided Erdelyi-Kober fractional integral of the order α>0 with respect to time variable [27, p. 105], (see also [83, p. 251], where β=σ, γ=η, δ=α). For σ=1, equation (53) takes the form of the Kober fractional integral (51). For η=0 and σ=1, equation (53) can be expressed through the Riemann-Liouville fractional integral (27) by the equation $(I_{0+;1,0}^{\alpha} X)(t) = t^{-\alpha} \cdot (I_{0+}^{\alpha} X)(t)$. Equations (51) and (53) describe generalized multiplier with memory.

Let us consider equation (53) for unit exogenous variable X(t)=1. Using equation (53) with X(t)=1 and equation 2.2.4.8 of [84, p. 296] in the form

$$\int_0^t \tau^{\sigma(\eta+1)-1} (t^{\sigma} - \tau^{\sigma})^{\alpha-1} d\tau = \sigma^{-1} \cdot t^{\sigma(\alpha+\eta)} \cdot B(\eta+1, \alpha), \qquad (54)$$

where $B(x,y) := \Gamma(x) \cdot \Gamma(y)/\Gamma(x+y)$ is the beta function (the Euler integral of first kind), we get

$$Y(t) = m \cdot (I_{0+;\sigma,\eta}^{\alpha} 1)(t) = \frac{m}{\Gamma(\alpha)} \cdot B(\eta+1, \alpha) = \frac{m \cdot \Gamma(\eta+1)}{\Gamma(\eta+\alpha+1)}. \qquad (55)$$

As a result, we get that the dynamic memory, which is defined by memory function (52), can be considered as a unit preserving memory up to a constant factor. This mean that processes with memory, which is described by equation (53), can remember a constant value of exogenous variables during the infinitely long period of time. Note that the memory (53) cannot be considered as a memory, which is homogeneous on time M(t,τ)≠M(t–τ) in general.

It should be noted that the memory function (52) is fading memory of power-law type. Using (52) and X(τ)=δ(τ–T), we get the expression

$$Y(t) = \int_0^t M_\sigma^{\alpha,\eta}(t,\tau) \cdot \delta(T-\tau) d\tau = \frac{\sigma \cdot m \cdot T^{\sigma(\eta+1)-1}}{\Gamma(\alpha)} \cdot t^{-\sigma(\alpha+\eta)} \cdot (t^\sigma - \tau^\sigma)^{\alpha-1}. \quad (56)$$

The Caputo modification of the Erdelyi-Kober fractional derivative has been suggested by Luchko and Trujillo [83, p. 260] and then it has been generalized by Kiryakova and Luchko in [85]. We can consider the expression

$$X_{new}(\tau) = \prod_{k=1}^n \left(\frac{\tau}{\sigma} \frac{d}{d\tau} + \eta + k\right) X(\tau), \quad (57)$$

where n-1<α≤n, instead of X(τ) and the memory function (52) in the form

$$M(t,\tau) = M_\sigma^{n-\alpha,\eta+\alpha}(t,\tau) = \frac{\sigma \cdot t^{-\sigma(n+\eta)}}{\Gamma(n-\alpha)} \frac{a \cdot \tau^{\sigma(\eta+\alpha+1)-1}}{(t^\sigma - \tau^\sigma)^{\alpha-n+1}}. \quad (58)$$

Substituting (57) and (58) into equation (5), we obtain the fractional differential equation

$$Y(t) = a \cdot (D_{0+;\sigma,\eta}^\alpha X)(t), \quad (59)$$

where $D_{0+;\sigma,\eta}^\alpha$ is the Caputo modification of the Erdelyi-Kober fractional derivative [83, p.260]. For η=0, σ=1, equation (59) can be expressed through the Caputo fractional derivative (39). Equation (59) represents the accelerator with memory of the generalized form.

### 4.7. Memory with power-law fading of distributed order

In economic processes with memory, the parameter α of memory fading can be distributed on the interval $[\alpha_1, \alpha_2]$, where the distribution is described by a weight function ρ(α). In this case, we should consider the memory function in the form, which depends on the weight function ρ(α) and the interval $[\alpha_1, \alpha_2]$.

For the memory with power-law fading of distributed order, we can use the memory function in the form

$$M(t-\tau) = M_{\rho(\alpha)}^{[\alpha_1,\alpha_2]}(t-\tau) = \int_{\alpha_1}^{\alpha_2} \frac{\rho(\alpha)}{\Gamma(\alpha)} \frac{m}{(t-\tau)^{1-\alpha}} d\alpha, \quad (60)$$

where $\alpha_2 > \alpha_1 \geq 0$ and the weight function ρ(α) satisfies the normalization condition

$$\int_{\alpha_1}^{\alpha_2} \rho(\alpha) d\alpha = 1. \quad (61)$$

Substituting (60) into equation (5), we obtain the integral equation in the form

$$Y(t) = m \left(I_{0+}^{[\alpha_1,\alpha_2]} X\right)(t), \quad (62)$$

where $I_{0+}^{[\alpha_1,\alpha_2]}$ is the Riemann-Liouville fractional integral of distributed order.

Using the left-sided Riemann-Liouville integral of the order α>0, which is defined by equation (27), the fractional integral of distributed order can be defined by the equation

$$\left(I_{0+}^{[\alpha_1,\alpha_2]}X\right)(t) := \int_{\alpha_1}^{\alpha_2} \rho(\alpha) \cdot (I_{0+}^{\alpha}X)(t)d\alpha, \tag{63}$$

where $\rho(\alpha)$ satisfies the normalization condition (61) and $\alpha_2 > \alpha_1 \geq 0$. In equation (63) the integration with respect to time and the integration with respect to order can be permuted for a wide class of functions $X(\tau)$. As a result, equation (63) can be represented in the form

$$\left(I_{0+}^{[\alpha_1,\alpha_2]}X\right)(t) := \int_0^t M_{\rho(\alpha)}^{[\alpha_1,\alpha_2]}(t-\tau) \cdot X(\tau)d\tau, \tag{64}$$

where the kernel $M_{\rho(\alpha)}^{[\alpha_1,\alpha_2]}(t-\tau)$ is defined by expression (60).

The concept of the integration and differentiation of distributed order was first proposed by Caputo in [86] and then developed in [87, 88, 89, 90]. Derivatives and integrals of distributed order are discussed in the book [91].

In equation (62), we can use the integer derivative $X^{(n)}(\tau) = d^n X(\tau)/d\tau^n$ as the exogenous variable. If $[\alpha_1] = [\alpha_2]$, we can use the memory function in the form

$$M_{\rho(n-\alpha)}^{[n-\alpha_2,n-\alpha_1]}(t-\tau) := \int_{\alpha_1}^{\alpha_2} \frac{\rho(\alpha)\cdot(t-\tau)^{n-\alpha-1}}{\Gamma(n-\alpha)} d\alpha = \int_{n-\alpha_2}^{n-\alpha_1} \frac{\rho(n-\alpha)\cdot(t-\tau)^{\alpha-1}}{\Gamma(\alpha)} d\alpha, \tag{65}$$

where $n := [\alpha_1] + 1 = [\alpha_2] + 1$. Substituting (65) into equation (6), where we use $X^{(n)}(\tau) = d^n X(\tau)/d\tau^n$, we obtain the fractional differential equation in the form

$$Y(t) = m\left(D_{0+}^{[\alpha_1,\alpha_2]}X\right)(t), \tag{66}$$

where $D_{0+}^{[\alpha_1,\alpha_2]}$ is the Caputo fractional derivative of distributed order.

Using the left-sided Caputo derivative of the order α>0, which is defined by equation (39), we can define the fractional derivative of distributed order in the form

$$\left(D_{0+}^{[\alpha_1,\alpha_2]}X\right)(t) := \int_{\alpha_1}^{\alpha_2} \rho(\alpha) \cdot (D_{0+}^{\alpha}X)(t)d\alpha, \tag{67}$$

where $n = [\alpha_1] + 1 = [\alpha_2] + 1$. Equation (67) can be represented in the form

$$\left(D_{0+}^{[\alpha_1,\alpha_2]}X\right)(t) := \int_0^t M_{\rho(n-\alpha)}^{[n-\alpha_2,n-\alpha_1]}(t-\tau) \cdot X^{(n)}(\tau)d\tau, \tag{68}$$

where the kernel $M_{\rho(n-\alpha)}^{[n-\alpha_2,n-\alpha_1]}(t-\tau)$ is defined by expression (65). Fractional derivatives of distributed order can be generalized for the case $[\alpha_1] < [\alpha_2]$.

In the simplest case, we can use the continuous uniform distribution (CUD) that is defined by the expression

$$\rho(\alpha) = \frac{1}{\alpha_2 - \alpha_1} \tag{69}$$

for the case $\alpha \in [\alpha_1, \alpha_2]$ with $\alpha_2 - \alpha_1 > 0$, and $\rho(\alpha)=0$ for other cases. For (69) the fractional derivative of distributed order is called the derivative of continual order or the continual derivative [92, 93, 94]. For the distribution function (69), the memory function has the form

$$M_{CUD}^{[\alpha_1,\alpha_2]}(t-\tau) = \frac{m}{\alpha_2-\alpha_1} \int_{\alpha_1}^{\alpha_2} \frac{(t-\tau)^{\alpha-1}}{\Gamma(\alpha)} d\alpha. \tag{70}$$

This function can be written in terms of the function

$$Vi(\alpha_1, \alpha_2, t) = \int_{\alpha_1}^{\alpha_2} \frac{t^\alpha}{\Gamma(\alpha)} d\alpha, \tag{71}$$

where $\alpha_2 \geq \alpha_1 \geq 0$ and $t \geq 0$, which has been suggested by Nahushev [92, p. 44]. Using function (71), expression (70) is represented in the form

$$M_{CUD}^{[\alpha_1,\alpha_2]}(t) = \frac{m}{(\alpha_2-\alpha_1) \cdot t} Vi(\alpha_1, \alpha_2, t). \tag{72}$$

The fractional derivatives of uniform distributed order (continual order) and its properties are described in the works [92, 93, 94].

In the general case, we can consider different distribution functions. These functions describe distributions of the parameter of memory fading on the set of economic agents. This is important for the economics, since various types of economic agents may have different parameters of memory fading.

In our opinion, the normal (or Gaussian) distribution can be of great interest for economic models. The normal (or Gaussian) distribution has the form

$$\rho(\alpha, \sigma, \mu) = \frac{1}{\sqrt{2\pi\sigma^2}} e^{-\frac{(\alpha-\mu)^2}{2\sigma^2}}, \tag{73}$$

where $\sigma$ is standard deviation, $\sigma^2$ is variance, $\mu$ is mean or expectation of the distribution. We can consider the case $0 \leq \alpha_1 < \mu < \alpha_2$. In this case, we have

$$\int_{\alpha_1}^{\alpha_2} \rho(\alpha, \sigma, \mu) d\alpha = N[\alpha_1, \alpha_2]. \tag{74}$$

Therefore we will use the distribution $\rho_{[\alpha_1,\alpha_2]}(\alpha,\sigma,\mu) = \rho(\alpha,\sigma,\mu)/N[\alpha_1,\alpha_2]$ to have the normalization condition in the form (61). One of the interesting case is the integer value of mean ($\mu=n$). In application, the normal distribution with integer mean (for example, $\mu=1$) and small variance can be used to describe wide class of economic processes with "weak" memory.

Using the left-sided Riemann-Liouville integral of the order $\alpha>0$, which is defined by equation (27), the fractional integral of the normal distributed order can be defined by the equation

$$\left(I_{0+,\mu,\sigma}^{[\alpha_1,\alpha_2]} X\right)(t) := \int_{\alpha_1}^{\alpha_2} \frac{\rho(\alpha,\sigma,\mu)}{N[\alpha_1,\alpha_2]} \cdot (I_{0+}^\alpha X)(t) d\alpha = \int_0^t M_{ND,\mu,\sigma}^{[\alpha_1,\alpha_2]}(t-\tau) \cdot X(\tau) d\tau, \tag{75}$$

where $M_{ND,\mu,\sigma}^{[\alpha_1,\alpha_2]}(t-\tau)$ is the memory function for the normal distribution (ND) such that

$$M_{ND,\mu,\sigma}^{[\alpha_1,\alpha_2]}(t-\tau) := \frac{m}{N[\alpha_1,\alpha_2]} \frac{1}{\sqrt{2\pi\sigma^2}} \int_{\alpha_1}^{\alpha_2} \frac{(t-\tau)^{\alpha-1}}{\Gamma(\alpha)} e^{-\frac{(\alpha-\mu)^2}{2\sigma^2}} d\alpha, \tag{76}$$

where $0 \leq \alpha_1 < \mu < \alpha_2$. Using the representation of the Dirac delta-function as a limit of the normal distribution ($\rho(\alpha, \sigma, \mu) \to \delta(\alpha - \mu)$ at $\sigma \to 0$), we get

$$\lim_{\sigma \to 0} M_{ND,\mu,\sigma}^{[\alpha_1,\alpha_2]}(t - \tau) = \frac{m}{N[\alpha_1,\alpha_2]} \frac{(t-\tau)^{\mu-1}}{\Gamma(\mu)}, \tag{77}$$

which is represented through the memory function (25) with $\alpha=\mu$. In this case, integral (75) is written through the Riemann-Liouville fractional integral (27).

Using the left-sided Caputo derivative of the order $\alpha>0$, which is defined by equation (39), we can define the fractional derivative of normal distributed order in the form

$$\left(D_{0+,\mu,\sigma}^{[\alpha_1,\alpha_2]}X\right)(t) := \int_{\alpha_1}^{\alpha_2} \frac{\rho(\alpha,\sigma,\mu)}{N[\alpha_1,\alpha_2]} \cdot (D_{0+}^{\alpha}X)(t) d\alpha, \tag{78}$$

where $\rho(\alpha, \sigma, \mu)$ is defined by equation (73) and $0 \leq \alpha_1 < \mu < \alpha_2$. For example, we can consider derivative (78) for the interval [0,2] and $\mu=1$ in the form

$$\left(D_{0+,\mu,\sigma}^{[0,2]}X\right)(t) := \int_0^2 \frac{\rho(\alpha,\sigma,1)}{N[0,2]} \cdot (D_{0+}^{\alpha}X)(t) d\alpha, \tag{79}$$

in which the order $\alpha$ is distributed near first order. Equation (79) can be represented in the form

$$\left(D_{0+,\mu,\sigma}^{[0,2]}X\right)(t) := \int_0^t M_{ND1,1,\sigma}^{[0,1]}(t - \tau) \cdot X^{(1)}(\tau) d\tau + \int_0^t M_{ND2,1,\sigma}^{[1,2]}(t - \tau) \cdot X^{(2)}(\tau) d\tau, \tag{80}$$

where the kernel $M_{NDn,\mu,\sigma}^{[n-1,n]}(t - \tau)$ with $\mu=1$ is defined by the equation

$$M_{NDn,1,\sigma}^{[n-1,n]}(t - \tau) = \frac{1}{N[0,2]} \frac{1}{\sqrt{2\pi\sigma^2}} \int_{n-1}^{n} \frac{(t-\tau)^{n-\alpha-1}}{\Gamma(n-\alpha)} e^{-\frac{(\alpha-1)^2}{2\sigma^2}} d\alpha. \tag{81}$$

For $\mu=1$, equation (78) with $0 \leq \alpha_1 < 1 < \alpha_2$ at the limit $\sigma \to 0$ gives the standard derivative of first order, i.e. $\left(D_{0+,1,0}^{[\alpha_1,\alpha_2]}X\right)(t) = X^{(1)}(t)$. The processes with "weak" memory, which is small deviation from classical case of amnesia, can be represented by the Gaussian distribution functions with integer mean and infinitesimally small variance. It is known that the delta-function can be considered as a limit of a family of the Gaussian functions, when the variance becomes smaller ($\rho(\alpha, \sigma, \mu) \to \delta(\alpha - \mu)$ at $\sigma \to 0$). The processes without memory are described by the Dirac delta-function $\rho(\alpha)=\delta(\alpha-n)$. Therefore the Gaussian distribution functions with integer mean and infinitesimally small variance can be used to describe economic processes with memory, which is distributed around the classical case.

## 5. Example of application of dynamic memory concept in macroeconomic: Harrod-Domar model with dynamic memory

In this section, we consider a macroeconomic model, in which we take into account memory effects with power-law fading. For simplification, we assume that the fading of dynamic memory has a power-like form, which is characterized by one parameter. We give solutions of the fractional differential equations that describe this macroeconomic model with power-law fading memory.

Using the asymptotic behavior of these solutions, we formulate principles of changing of technological growth rates for macroeconomic dynamics with power-law memory.

One of the simplest models of economic growth is the Harrod-Domar model [77], which combines the results of Harrod [98] and Domar [99, 100]. The Harrod-Domar continuous time model describes the dynamics of national income Y(t), which is determined by the sum of the non-productive consumption C(t) and the investment I(t) (accumulation of basic production assets). Therefore, the balance equation of this model has the form

$$Y(t) = I(t) + C(t). \tag{82}$$

In the Harrod-Domar model, the non-productive consumption C(t) is considered as a function independent of income and investment.

The standard Harrod–Domar model is usually formulated by using the following assumptions: (a) the proportionality of production accumulation and the growth of the national income at the same time; (b) an instantaneous transformation of investment (capital investment) in the growth of national income, which means that the memory effects are neglected; (c) the independence of consumption dynamics. The first assumption gives a relationship between investment and the growth of the national income. This relationship is represented by the accelerator equations that describes a direct proportionality between investment (productive accumulation) and the growth of the national income in the form

$$I(t) = B \cdot \frac{dY(t)}{dt}, \tag{83}$$

where B is the accelerator coefficient, which describes the capital intensity of the national income (incremental capital intensity, differential capital intensity). Substituting expression (83) into equation (82), we obtain

$$Y(t) = B \cdot \frac{dY(t)}{dt} + C(t). \tag{84}$$

Equation (84) describes the dynamics of the national income within framework of the Harrod–Domar model. Equation (84) shows that the dynamics of the national income Y(t) is determined by the behavior of the function C(t) if the parameter B is given.

The derivatives of the first order, which are used in equations (83) and (84), imply an instantaneous change of the investment I(t), when changing the growth rate of the national income Y(t). In other words, the investments at time t are determined by the properties of the national income in an infinitesimal neighborhood of this time point. Because of this, accelerator equation (83) does not take into account the effects of dynamic memory. As a result, differential equation (84) can be used only to describe an economy, in which all economic agents have an instantaneous amnesia. This restriction substantially narrows the field of application of macroeconomic models to describe the real economic processes. In many cases, economic agents can remember the history of changes of the gross product, national income and investment. As a result, this memory about history of changes of investment and income can influence on decision-making by economic agents.

To take into account the effects of power-law memory, we can apply the mathematical tools of derivatives of non-integer orders. In the case of dynamic memory with one-parametric power-law fading, the linear equation of the accelerator with memory is written in the form

$$I(t) = B \cdot (D_{0+}^{\alpha} Y)(t), \tag{85}$$

where $(D_{0+}^{\alpha} Y)(t)$ is the Caputo derivative of the order $\alpha \geq 0$ that is defined by equation (39). In general, the capital intensity depends on the parameter of memory fading, i.e. $B=B(\alpha)$. For $\alpha=1$ equation (85) gives equation (83), since the Caputo fractional derivative of the order $\alpha=1$ coincides with the first derivative.

Substituting the expression for the investment $I(t)$, which is given by formula (85), into balance equation (82), we obtain the fractional differential equation

$$Y(t) = B \cdot (D_{0+}^{\alpha} Y)(t) + C(t). \tag{86}$$

For $\alpha=1$ equation (86) gives equation (84).

Equation (86) determines the dynamics of the national income within the framework of the proposed macroeconomic model with one-parametric power-law memory. If the parameter B is given, then the dynamics of national income $Y(t)$ is determined by the behavior of the function $C(t)$. In the Harrod-Domar model, the non-productive consumption $C(t)$ is assumed independent of income and investment. If non-productive consumption is represented as a fixed part of income $C(t) = c \cdot Y(t)$, where c is the marginal propensity to consume, then equation (86) describes the natural growth model with memory [68, 69, 70, 71].

We can solve equation (86) by using Theorem 5.15 of book [27, p. 323]. Fractional differential equation (86) can be written in the form

$$(D_{0+}^{\alpha} Y)(t) - B^{-1} \cdot Y(t) = -B^{-1} \cdot C(t), \tag{87}$$

where $n - 1 < \alpha \leq n$. If C (t) is a continuous function defined on the positive semiaxis (t>0), then equation (87) has the solution

$$Y(t) = Y_C(t) + \sum_{k=0}^{n-1} Y^{(k)}(0) \cdot t^k \cdot E_{\alpha, k+1}[B^{-1} \cdot t^{\alpha}], \tag{88}$$

where $Y^{(k)}(0)$ is derivatives of the integer order $k \geq 0$ of the function $Y(t)$ at $t=0$, and

$$Y_C(t) := -B^{-1} \cdot \int_0^t (t - \tau)^{\alpha - 1} \cdot E_{\alpha, \alpha}[B^{-1} \cdot (t - \tau)^{\alpha}] \cdot C(\tau) d\tau. \tag{89}$$

Here $E_{\alpha, \beta}[z]$ is the two-parametric Mittag-Leffler function [101] that is defined by the expression

$$E_{\alpha, \beta}[z] := \sum_{k=0}^{\infty} \frac{z^k}{\Gamma(\alpha k + \beta)}, \tag{90}$$

where $\alpha > 0$, and $\beta$ is an arbitrary real or complex number.

For $0 < \alpha \leq 1$ (n=1) solution (88) with (89) of equation (87) has the form

$$Y(t) = -B^{-1} \cdot \int_0^t (t-\tau)^{\alpha-1} \cdot E_{\alpha,\alpha}[B^{-1} \cdot (t-\tau)^\alpha] \cdot C(\tau) d\tau + Y(0) \cdot E_{\alpha,1}[B^{-1} \cdot t^\alpha]. \quad (91)$$

In the generalized Harrod–Domar model, which is described by equation (87), the non-productive consumption C(t) is considered as a function, which is independent of national income and investment. Therefore this function can be a constant in time, a power-law function, or it can be a more complicated function. Let us consider the consumption function in the form

$$C(t) = C \cdot t^{\mu-1}, \quad (92)$$

where $\mu > 0$. The constant consumption function (C(t)=C) is a special case of (92), when $\mu = 1$.

Using formula (4.4.5) of [101, p. 61], we get

$$Y_C(t) = -C \cdot B^{-1} \cdot \Gamma(\mu) \cdot t^{\alpha+\mu-1} \cdot E_{\alpha,\alpha+\mu}[B^{-1} \cdot t^\alpha], \quad (93)$$

where $\mu > 0$. As a result, the general solution of equation (87) with function (92) has the form

$$Y(t) = \sum_{k=0}^{n-1} Y^{(k)}(0) \cdot t^k \cdot E_{\alpha,k+1}[B^{-1} \cdot t^\alpha] - C \cdot B^{-1} \cdot \Gamma(\mu) \cdot t^{\alpha+\mu-1} \cdot E_{\alpha,\alpha+\mu}[B^{-1} \cdot t^\alpha], \quad (94)$$

where $0 < n-1 < \alpha \leq n$.

Using formula (4.2.3) of [101, p. 57], expression (93) can be written in the form

$$Y_C(t) = C \cdot t^{\mu-1} \cdot \left(1 - \Gamma(\mu) \cdot E_{\alpha,\mu}[B^{-1} \cdot t^\alpha]\right). \quad (95)$$

Therefore the equation of the Harrod–Domar model with power-law memory, whose fading parameter is α ($0 < n-1 < \alpha \leq n$) and the consumption function (92), has the solution

$$Y(t) = C \cdot t^{\mu-1} \cdot \left(1 - \Gamma(\mu) \cdot E_{\alpha,\mu}[B^{-1} \cdot t^\alpha]\right) + \sum_{k=0}^{n-1} Y^{(k)}(0) \cdot t^k \cdot E_{\alpha,k+1}[B^{-1} \cdot t^\alpha]. \quad (96)$$

For μ=1, equation (96) describes the case of a constant consumption function (C(t)=C), where the solution has the form

$$Y(t) = C \cdot (1 - E_{\alpha,1}[B^{-1} \cdot t^\alpha]) + \sum_{k=0}^{n-1} Y^{(k)}(0) \cdot t^k \cdot E_{\alpha,k+1}[B^{-1} \cdot t^\alpha]. \quad (97)$$

Solution (97) describes economic growth with one-parametric power-law memory about changes in income and investment with a constant non-productive consumption. For $0 < \alpha \leq 1$ (n=1) solution (97) takes the form

$$Y(t) = C \cdot (1 - E_{\alpha,1}[B^{-1} \cdot t^\alpha] - 1) + Y(0) \cdot E_{\alpha,1}[B^{-1} \cdot t^\alpha]. \quad (98)$$

For α=1, the equality $E_{\alpha,1}[z] = e^z$, leads solution (98) to the form

$$Y(t) = C \cdot (1 - \exp(B^{-1} \cdot t)) + Y(0) \cdot \exp(B^{-1} \cdot t), \quad (99)$$

which exactly coincides with solution of the standard model (84). Solution (99) of equation (87) describes economic growth in the framework of standard macroeconomic model with constant consumption that does not take into account memory effects.

A macroeconomic model is called closed if it assumes the absence of non-productive consumption (C(t)=0). The assumption C(t)=0 is unrealistic, but the consideration of this case allows us to estimate the greatest possible growth rate of national income, which is limited only by the incremental capital intensity. The solution (88) with (89) of equation (87) with C(t)=0 has the form

$$Y(t) = \sum_{k=0}^{n-1} Y^{(k)}(0) \cdot t^k \cdot E_{\alpha,k+1}[B^{-1} \cdot t^\alpha]. \tag{100}$$

For 0<α<1, solution (100) takes the form $Y(t) = Y(0) \cdot E_{\alpha,1}[B^{-1} \cdot t^\alpha]$. For α=1, solution (100) gives expression $Y(t) = Y(0) \cdot \exp(t/B)$, where the value $\lambda = B^{-1}$ describes the technological growth rate, which does not take into account the memory effects.

Let us describe the behavior of solution (100) at $t \to \infty$ in the framework of the Harrod-Domar model with one-parametric power-law memory. Using equation (1.8.27) of book [27, p. 43] for 0<α<2, we get the equation

$$E_{\alpha,\beta+1}[\lambda \cdot t^\alpha] = \frac{\lambda^{-\beta/\alpha}}{\alpha} \cdot t^{-\beta} \cdot \exp(\lambda^{1/\alpha} \cdot t) - \sum_{j=1}^{m} \frac{\lambda^{-j}}{\Gamma(\beta+1-\alpha \cdot j)} \cdot \frac{1}{t^{\alpha \cdot j}} + O\left(\frac{1}{t^{\alpha \cdot (m+1)}}\right) \tag{101}$$

for $t \to \infty$, where $\lambda = B^{-1}$ is a real number. Using equation (1.8.29) of book [27, p. 43] for α≥2 and $t \to \infty$, we get the asymptotic expression

$$E_{\alpha,\beta+1}[\lambda \cdot t^\alpha] = \frac{\lambda^{-\beta/\alpha}}{\alpha} \cdot t^{-\beta} \cdot \sum_N \exp\left(\lambda^{1/\alpha} \cdot t \cdot \exp\left(i\frac{2\pi \cdot N}{\alpha}\right)\right) \cdot \exp\left(i\frac{2\pi \cdot k \cdot N}{\alpha}\right) -$$
$$\sum_{k=1}^{m} \frac{\lambda^{-k}}{\Gamma(\beta+1-\alpha \cdot k)} \cdot \frac{1}{t^{\alpha \cdot k}} + O\left(\frac{1}{t^{\alpha \cdot (m+1)}}\right), \tag{102}$$

where $\lambda = B^{-1}$ is a real number ($\arg(\lambda) = 0$) and the first sum is taken over all integer values of N that satisfy the condition $|N| \leq \alpha/4$. Using equation (102) and the inequality $|N| \leq \alpha/4$, we see that equation (101) can be used for α<4 in the considered model.

Asymptotic equation (101) allows us to describe the behavior at $t \to \infty$ in a macroeconomic model with power-law memory, in which the memory fading parameter is 0<α<2 (and 0<α<4). Substitution of expression (101) with $\lambda = B^{-1}$ and β=k into (100) gives

$$Y(t) = \sum_{k=0}^{n-1} Y^{(k)}(0) \cdot \frac{B^{k/\alpha}}{\alpha} \cdot \exp(B^{-1/\alpha} \cdot t) + \sum_{k=0}^{n-1}\left(\sum_{j=1}^{m} \frac{Y^{(k)}(0) \cdot B^j}{\Gamma(k+1-\alpha \cdot j)} \cdot t^{k-\alpha \cdot j} + O\left(\frac{1}{t^{\alpha \cdot (m+1)-k}}\right)\right), \tag{103}$$

where 0<n-1<α<n. Expression (103) describes behavior of solution (100) at $t \to \infty$.

As a result, we find that the technological growth rates of Harrod-Domar model with one-parametric memory do not coincide with the growth rates $\lambda = B^{-1}$ of standard Harrod-Domar models without memory. The technological growth rate with memory is equal to the value

$$\lambda_{\text{eff}}(\alpha) := \lambda^{1/\alpha} = B^{-1/\alpha}, \tag{104}$$

which will be called effective technological growth rate of the model with one-parametric power-law memory.

For open macroeconomic model (C(t)≠0). To describe behavior of solution (93) at $t \to \infty$, we can use expression (101). Substitution of (101) with $\lambda = B^{-1}$ and $\beta = \alpha + \mu - 1$ into (93) gives

$$Y_C(t) = -\frac{C \cdot \Gamma(\mu) \cdot B^{(1-2\alpha-\mu)/\alpha}}{\alpha} \cdot \exp(B^{-1/\alpha} \cdot t) + \sum_{j=1}^{m} \frac{C \cdot \Gamma(\mu) \cdot B^{j-1}}{\Gamma(\alpha(1-j)+\mu)} \cdot t^{\alpha(1-j)+\mu-1} + O\left(\frac{1}{t^{\alpha \cdot m+1-\mu}}\right) \tag{105}$$

where μ > 0. As a result, we can see that the technological growth rate of solution (94), which described the macroeconomic model with one-parametric memory, is defined by equation (104), i.e. $\lambda_{eff}(\alpha) = B^{-1/\alpha}$.

Using the suggested equation, we can compare the technological growth rates in the macroeconomic model with one-parametric power-law memory and the rates of the standard model, which does not take into account the memory effects. We can see that the account of memory effects can significantly change the technological growth rates in the macroeconomic model. At the same time, the technological growth rates may both increase and decrease in comparison with the standard model, which does not take into account the memory effects. If the incremental capital intensity of the national income is less than one (0 < B < 1), then the technological growth rate of the economy can be much larger, when we consider the memory effects. For example, for α=0.2 and B= 0.5, the technological growth rate of the model with memory is equal to $\lambda_{eff}(0.5) = 32$ instead of λ = 2 for the standard model, that is, 16 times more.

The obtained results allow us to formulate the following principles of macroeconomic dynamics, which take into account the effects of one-parametric power-law memory.

**Principles of changing of technological growth rates:**

*I) In Harrod-Domar model, the effects of one-parametric power-law memory with the fading parameter α>0, lead to the dependence of the technological growth rates on the parameter α, which are determined by the formula*

$$\lambda_{eff}(\alpha) = B^{-1/\alpha}, \tag{106}$$

*where B is the incremental capital intensity that characterizes the amount of investment corresponding to the change of the growth rate of the national income.*

*II) For small technological growth rates, which are described by the standard Harrod-Domar model, the effects of one-parameter memory with the fading parameter $0 < \alpha < 1$ lead to decrease of the growth rates of the economy, and its lead us to an increase of the growth rates for $\alpha > 1$.*

*III) For the large rates of technological growth, which are described by the standard Harrod-Domar model, the effects of power-law memory with the fading $0 < \alpha < 1$ leads to an increase of the growth rates of the economy, and its lead to a decrease of growth rates for $\alpha > 1$.*

We see that neglecting the memory effects in macroeconomic models can greatly change the result. Accounting of the memory effects can lead to new results for the same parameters of macroeconomic models. We demonstrate that the memory effects can change the economic growth rate and change dominant parameters, which determine growth rates. The principle of changing of technological growth rates and the principle of domination change have been suggested in [102] for the intersectoral economic dynamics with power-law memory. It has been shown that in the input–output economic dynamics the effects of fading memory can change the economic growth rate and dominant behavior of economic sectors. Fractional dynamics of sectors of national economies, where the parameters depend on time, are discussed in [103]. Equations of dynamic intersectoral model with power-law memory, where the matrix of direct material costs and the matrix of incremental capital intensity of production depend on time, and the solutions of these equations are suggested in [103]. Economic processes with power-law memory are also considered in papers [57-75], where generalizations of some basic economic concepts have been proposed.

## 6. Conclusion

The use of the concept of dynamic memory, which is described in this paper, allows us to build mathematical models of economic processes with different types of memory. The use of differential equations with derivatives of non-integer orders can allow us to obtain solutions by using the methods of the fractional calculus [25, 26, 27, 28, 82, 92, 94, 104, 105]. For some models of economic processes we can derive analytical solutions. The use of numerical methods [95, 96, 97] of the fractional calculus can used to realize computer simulations of economic processes with dynamic memory.

In economic models we should take into account the memory effects that are caused by the fact that economic agents remember the story of changes of exogenous and endogenous variables that characterize the economic process. The agents can take into account these changes in making economic decisions. The continuous time description of the economic processes with the power-law fading memory can be based on the fractional calculus and the fractional differential equations. The inclusion of memory effects into the economic models can lead to qualitatively new results at the same the parameters. In constructions of appropriate economic models we should take into account a possible dependence of economic processes on memory effects to predict the dynamics of economy. Therefore, it is important to develop the concept of dynamic memory to describe real economic processes.

**References**

1. Volterra V. Theory of functionals and of integral and integro-differential equations. Dover, 2005. 299 p. Chapter VI, Section IV.
2. Wang C.C. The principle of fading memory // Archive for Rational Mechanics and Analysis. 1965. Vol. 18. No. 5. P. 343–366. DOI: 10.1007/BF00281325
3. Coleman B.D., Mizel V.J. On the general theory of fading memory // Archive for Rational Mechanics and Analysis. 1968. Vol.29. No. 1. P. 18–31. DOI: 10.1007/BF00256456

4. Coleman B.D., Mizel V.J. A general theory of dissipation in materials with memory // Archive for Rational Mechanics and Analysis. 1967. Vol. 27. No. 4. P. 255–274. DOI: 10.1007/BF00281714
5. Coleman B.D., Mizel V.J. Norms and semi-groups in the theory of fading memory // Archive for Rational Mechanics and Analysis. 1966. Vol. 23. No. 2. P. 87–123. DOI: 10.1007/BF00251727
6. Saut J.C., Joseph D.D. Fading memory // Archive for Rational Mechanics and Analysis. 1983. Vol. 81. No. 1. P. 53–95. DOI: 10.1007/BF00283167
7. Day W.A. The Thermodynamics of Simple Materials with Fading Memory. Berlin: Springer-Verlag, 1972. 134 p.
8. Rabotnov Yu.N. Elements of Hereditary Solid Mechanics. Moscow: Mir Publishers, 1980. 387 p.
9. Lokshin A.A., Suvorova Yu.V. Mathematical Theory of Wave Propagation in Media with Memory. Moscow: Moscow State University. 1982. 152 p. [in Russian]
10. Alber H.D. Materials with Memory. Initial Boundary Value Problems for Constitutive Equations with Internal Variables. Berlin: Springer-Verlag, 1998. 171 p.
11. Mainardi F. Fractional Calculus and Waves Linear Viscoelasticity: An Introduction to Mathematical Models. London: Imperial College Press, 2010. 368 p.
12. Tarasov V.E. Fractional Dynamics: Applications of Fractional Calculus to Dynamics of Particles, Fields and Media. New York: Springer, 2010. 505 p. DOI: 10.1007/978-3-642-14003-7
13. Amendola G., Fabrizio M., Golden J.M. Thermodynamics of Materials with Memory: Theory and Applications. Springer Science & Business Media, 2011. 576 p. DOI: 10.1007/978-1-4614-1692-0
14. Teyssiere G., Kirman A.P. Long Memory in Economics. Berlin, Heidelberg: Springer-Verlag, 2007. 390 p.
15. Baillie R.N. Long memory processes and fractional integration in econometrics // Journal of Econometrics. 1996. Vol. 73. P. 5-59. DOI: 10.1016/0304-4076(95)01732-1
16. Banerjee A., Urga G. Modelling structural breaks, long memory and stock market volatility: an overview // Journal of Econometrics. 2005. Vol. 129. No. 1-2. P. 1-34. DOI: 10.1016/j.jeconom.2004.09.001
17. Beran, J. Statistics for Long-Memory Processes. New York: Capman and Hall, 1994. 315 p.
18. Beran J., Feng Y., Ghosh S. Kulik R. Long-Memory Processes: Probabilistic: Properties and Statistical Methods. Berlin: Springer-Verlag, 2013. DOI: 10.1007/978-3-642-35512-7
19. Time Series with Long Memory: Advanced Texts in Econometrics. Edited by P.M. Robinson. Oxford: Oxford University Press, 2003. 382 p.
20. Granger C.W.J., Joyeux R. An introduction to long memory time series models and fractional differencing // Journal of Time Series Analysis. 1980. Vol. 1. P. 15-39. DOI: 10.1111/j.1467-9892.1980.tb00297.x
21. Hosking J.R.M. Fractional differencing // Biometrika. 1981. Vol. 68. No. 1. P. 165-176. DOI: 10.1093/biomet/68.1.165
22. Parke W.R. What is fractional integration? // Review of Economics and Statistics. 1999. Vol. 81. No. 4. P. 632-638. DOI: 10.1162/003465399558490

23. Ghysels E., Swanson N.R., Watson M.W. Essays in Econometrics Collected Papers of Clive W.J. Granger. Volume II: Causality, Integration and Cointegration, and Long Memory. Cambridge: Cambridge University Press, 2001. 398 p.
24. Gil-Alana L.A., Hualde J. Fractional integration and cointegration: an overview and an empirical application. Palgrave Handbook of Econometrics. Volume 2: Applied Econometrics. Edited by T.C. Mills, K. Patterson. Berlin: Springer-Verlag, 2009. pp. 434-469. DOI: 10.1057/9780230244405_10
25. Samko S.G., Kilbas A.A., Marichev O.I. Fractional Integrals and Derivatives Theory and Applications. New York: Gordon and Breach, 1993. 1006 p.
26. Podlubny I. Fractional Differential Equations. San Diego: Academic Press, 1998. 340 p.
27. Kilbas A.A., Srivastava H.M., Trujillo J.J. Theory and Applications of Fractional Differential Equations. Amsterdam: Elsevier, 2006. 540 p.
28. Diethelm K. The Analysis of Fractional Differential Equations: An Application-Oriented Exposition Using Differential Operators of Caputo Type. Berlin: Springer-Verlag, 2010. 247 p. DOI: 10.1007/978-3-642-14574-2
29. Tarasov V.E., Tarasova V.V. Long and short memory in economics: fractional-order difference and differentiation // IRA-International Journal of Management and Social Sciences. 2016. Vol. 5. No. 2. P. 327-334. DOI: 10.21013/jmss.v5.n2.p10
30. Skovranek T., Podlubny I., Petras I. Modeling of the national economies in state-space: A fractional calculus approach // Economic Modelling. 2012. Vol. 29. No. 4. P. 1322–1327. 10.1016/j.econmod.2012.03.019
31. Tenreiro Machado J.A., Mata M.E. Pseudo phase plane and fractional calculus modeling of western global economic downturn // Communications in Nonlinear Science and Numerical Simulation. 2015. Vol. 22. No. 1-3. P. 396–406. DOI: 10.1016/j.cnsns.2014.08.032
32. Tenreiro Machado J.A., Mata M.E. Lopes A.M. Fractional state space analysis of economic systems // Entropy. 2015. Vol. 17. No. 8. P. 5402-5421. DOI: 10.3390/e17085402
33. Tejado I., Valerio D., Valerio N. Fractional calculus in economic growth modelling. The Spanish case // CONTROLO'2014 – Proceedings of the 11th Portuguese Conference on Automatic Control. Volume 321 of the series Lecture Notes in Electrical Engineering. Edited by A.P. Moreira, A. Matos, G. Veiga Springer International Publishing, 2015. P. 449-458. DOI: 10.1007/978-3-319-10380-8_43
34. Tejado I., Valerio D., Perez E., Valerio N. Fractional calculus in economic growth modelling: the Spanish and Portuguese cases // International Journal of Dynamics and Control. 2017. Vol. 5. No. 1. P. 208-222. DOI: 10.1007/s40435-015-0219-5
35. Tejado I., Valerio D., Perez E., Valerio N. Fractional calculus in economic growth modelling: The economies of France and Italy // Proceedings of International Conference on Fractional Differentiation and its Applications. Novi Sad, Serbia, July 18–20. Edited by D.T. Spasic, N. Grahovac, M. Zigic, M. Rapaic, T.M. Atanackovic. 2016. P. 113–123.
36. Scalas E., Gorenflo R., Mainardi F. Fractional calculus and continuous-time finance // Physica A. 2000. Vol. 284. No. 1–4. P. 376–384. DOI: 10.1016/S0378-4371(00)00255-7
37. Mainardi F., Raberto M., Gorenflo R., Scalas E. Fractional calculus and continuous-time finance II: The waiting-time distribution // Physica A. 2000. Vol. 287. No. 3–4. P. 468–481. DOI: 10.1016/S0378-4371(00)00386-1
38. Laskin N. Fractional market dynamics // Physica A. 2000. Vol. 287. No. 3. P. 482–492. DOI: 10.1016/S0378-4371(00)00387-3

39. Plerou V., Gopikrishnan P., Amaral L.A.N., Gabaix X., Stanley H.E. Economic fluctuations and anomalous diffusion // Physical Review E. 2000. Vol. 62. P. R3023–R3026. DOI: 10.1103/PhysRevE.62.R3023
40. Gorenflo R., Mainardi F., Raberto M., Scalas E. Fractional diffusion in finance: basic theory // MDEF2000. Workshop "Modelli Dinamici in Economia e Finanza", Urbino, Italy, September 28-30, 2000. Available at http://www.mdef.it/fileadmin/user_upload/mdef/meetings/MDEF2000/MainardiMDEF.pdf
41. Gorenflo R., Mainardi F., Scalas E., Raberto M. Fractional calculus and continuous-time finance III: the diffusion limit // in Mathematical Finance. Workshop of the Mathematical Finance Research Project, Konstanz, Germany, October 5–7, 2000. Edited by A. Kohlman, S. Tang. Basel: Birkhäuser Basel, 2001. P. 171–180. DOI: 10.1007/978-3-0348-8291-0_17
42. Raberto M., Scalas E., Mainardi F., Waiting-times and returns in high-frequency financial data: an empirical study // Physica A. 2002. Vol. 314. No.1-4. P. 749-755. DOI: 10.1016/S0378-4371(02)01048-8
43. West B.J., Picozzi S. Fractional Langevin model of memory in financial time series // Physical Review E. 2002. Vol. 65. Article ID 037106. DOI: 10.1103/PhysRevE.65. 037106
44. Picozzi S., West B.J. Fractional Langevin model of memory in financial markets // Physical Review E. 2002. Vol. 66. Article ID 046118. 12 p. DOI: 10.1103/PhysRevE.66.046118
45. Scalas E. Five years of continuous-time random walks in econophysics // Chapter in The Complex Networks of Economic Interactions. Volume 567. Lecture Notes in Economics and Mathematical Systems. Berlin: Springer-Verlag, Berlin 2006. P. 3-16. DOI: 10.1007/3-540-28727-2_1
46. Scalas E., The application of continuous-time random walks in finance and economics // Physica A. 2006. Vol. 362. No. 2. P. 225–239. DOI: 10.1016/j.physa.2005.11.024
47. Meerschaert M.M., Scalas E. Coupled continuous time random walks in finance // Physica A. 2006. Vol. 370. No. 1. P. 114-118. DOI: 10.1016/j.physa.2006.04.034
48. Cartea A., Del-Castillo-Negrete D. Fractional diffusion models of option prices in markets with jumps // Physica A. 2007. Vol. 374. No. 2. P. 749–763. DOI: 10.2139/ssrn.934809
49. Blackledge J. Application of the fractal market hypothesis for modelling macroeconomic time series // ISAST Trans Electron Signal Process. 2008. Vol. 2. No. 1. P. 89–110.
50. Mendes R.V. A fractional calculus interpretation of the fractional volatility model // Nonlinear Dynamics. 2009. Vol. 55. No. 4. P. 395–399. DOI: 10.1007/s11071-008-9372-0
51. Blackledge J. Application of the fractional diffusion equation for predicting market behavior // International Journal of Applied Mathematics. 2010. Vol. 40. No. 3. P. 130-158.
52. Tenreiro Machado J., Duarte F.B., Duarte G.M. Fractional dynamics in financial indices // International Journal of Bifurcation and Chaos. 2012. Vol. 22. No. 10. Article ID 1250249. 12 p. DOI: 10.1142/S0218127412502495
53. Kerss A., Leonenko N., Sikorskii A. Fractional Skellam processes with applications to finance // Fractional Calculus and Applied Analysis. 2014. Vol. 17. No. 2. P. 532-551. DOI: 10.2478/s13540-014-0184-2
54. Kleinerta H., Korbel J. Option pricing beyond Black-Scholes based on double-fractional diffusion // Physica A. 2016. Vol. 449. P. 200–214. DOI: 10.1016/j.physa.2015.12.125
55. Korbel J., Luchko Yu. Modeling of financial processes with a space-time fractional diffusion equation of varying order // Fractional Calculus and Applied Analysis. 2016. Vol. 19. No. 6. P. 1414–1433. DOI: 10.1515/fca-2016-0073

56. Fallahgoul H.A., Focardi S.M., Fabozzi F.J. Fractional Calculus and Fractional Processes with Applications to Financial Economics, Theory and Application. London: Academic Press, 2016. 118 p. ISBN: 9780128042489
57. Tarasova V.V., Tarasov V.E. Economic interpretation of fractional derivatives // Progress in Fractional Differentiation and Applications. 2017. Vol. 3. No. 1. P. 1-7. DOI: 10.18576/pfda/030101
58. Tarasova V.V., Tarasov V.E. Criteria hereditarity of economic process and memory effect // Young scientist [Molodoj Uchenyj]. 2016. No. 14 (118). P. 396–399. [in Russian]
59. Tarasova V.V., Tarasov V.E. Marginal utility for economic processes with memory // Almanac of Modern Science and Education [Almanah Sovremennoj Nauki i Obrazovaniya]. 2016. No. 7 (109). P. 108–113. [in Russian]
60. Tarasova V.V., Tarasov V.E. Marginal values of non-integer order in economic analysis // Azimuth Scientific Research: Economics and Management [Azimut Nauchnih Issledovanii: Ekonomika i Upravlenie]. 2016. No. 3 (16). P. 197-201. [in Russian]
61. Tarasova V.V., Tarasov V.E. Economic indicator that generalizes average and marginal values // Journal of Economy and Entrepreneurship [Ekonomika i Predprinimatelstvo]. 2016. No. 11-1 (76-1). P. 817-823. [in Russian]
62. Tarasova V.V., Tarasov V.E. A generalization of concepts of accelerator and multiplier to take into account memory effects in macroeconomics // Journal of Economy and Entrepreneurship [Ekonomika i Predprinimatelstvo]. 2016. No. 10-3 (75-3). P. 1121-1129. [in Russian]
63. Tarasova V.V., Tarasov V.E. Economic accelerator with memory: discrete time approach // Problems of Modern Science and Education [Problemy Sovremennoj Nauki i Obrazovaniya]. 2016. No. 36 (78). P. 37-42. DOI: 10.20861/2304-2338-2016-78-002
64. Tarasova V.V., Tarasov V.E. Elasticity for economic processes with memory: fractional differential calculus approach // Fractional Differential Calculus. 2016. Vol. 6. No. 2. P. 219-232. DOI: 10.7153/fdc-06-14
65. Tarasova V.V., Tarasov V.E. Measures of risk aversion for investors with memory: hereditary generalization of Arrow-Pratt measure // Financial Journal [Finansovyj Zhurnal]. 2017. No. 2 (36). P. 46-63. [in Russian]
66. Tarasova V.V., Tarasov V.E. Non-local measures of risk aversion in the economic process // Economics: Theory and Practice [Ekonomika: Teoriya i Praktika]. 2016. No. (44). P. 54-58. [in Russian]
67. Tarasova V.V., Tarasov V.E. Deterministic factor analysis: methods of integro-differentiation of non-integral order // Actual Problems of Economics and Law [Aktualnye Problemy Ekonomiki i Prava]. 2016. Vol. 10. No. 4. P. 77–87. [in Russian] DOI: 10.21202/1993-047X.10.2016.4.77-87
68. Tarasova V.V., Tarasov V.E. Influence of memory effects on world economics and business // Azimuth Scientific Research: Economics and Management [Azimut Nauchnih Issledovanii: Ekonomika i Upravlenie]. 2016. Vol. 5. No. 4 (17). P. 369-372. [in Russian]
69. Tarasova V.V., Tarasov V.E. Economic growth model with constant pace and dynamic memory // Problems of Modern Science and Education. 2017. No. 2 (84). P. 38-43. DOI: 10.20861/2304-2338-2017-84-001

70. Tarasova V.V., Tarasov V.E. Fractional dynamics of natural growth and memory effect in economics // European Research. 2016. No. 12 (23). P. 30-37. DOI: 10.20861/2410-2873-2016-23-004
71. Tarasova V.V., Tarasov V.E. Logistic map with memory from economic model // Chaos, Solitons and Fractals. 2017. Vol. 95. P. 84-91. DOI: 10.1016/j.chaos.2016.12.012
72. Tarasova V.V., Tarasov V.E. Hereditary generalization of Harrod-Domar model and memory effects // Journal of Economy and Entrepreneurship [Ekonomika i Predprinimatelstvo]. 2016. No. 10-2 (75-2). P. 72-78.
73. Tarasova V.V., Tarasov V.E. Memory effects in hereditary Harrod-Domar model // Problems of Modern Science and Education [Problemy Sovremennoj Nauki i Obrazovaniya]. 2016. No. 32 (74). P. 38-44. [in Russian] DOI: 10.20861/2304-2338-2016-74-002
74. Tarasova V.V., Tarasov V.E. Keynesian model of economic growth with memory // Economics and Management: Problems, Solutions [Ekonomika i Upravlenie: Problemy i Resheniya]. 2016. No. 10-2 (58). P. 21-29. [in Russian]
75. Tarasova V.V., Tarasov V.E. Memory effects in hereditary Keynes model // Problems of Modern Science and Education [Problemy Sovremennoj Nauki i Obrazovaniya]. 2016. No. 38 (80). P. 56-61. [in Russian] DOI: 10.20861/2304-2338-2016-80-001
76. Tarasova V.V., Tarasov V.E. Elasticity of OTC cash turnover of currency market of Russian Federation // Actual Problems of Humanities and Natural Sciences. [Aktualnye Problemy Gumanitarnyh i Estestvennyh Nauk]. 2016. No. 7–1 (90). P. 207–215. [in Russian]
77. Allen R.G.D. Mathematical Economics. Second edition. London: Macmillan, 1960. 812 p. DOI 10.1007/978-1-349-81547-0
78. Ross B., Samko S.G. Fractional integration operator of variable order in the Hölder spaces $H^{\lambda(x)}$ // International Journal of Mathematics and Mathematical Sciences. 1995. Vol. 18. No. 4. P. 777–788.
79. Samko S.G. Fractional integration and differentiation of variable order // Analysis Mathematica. 1995. Vol. 21. No. 3. P. 213–236. DOI: 10.1007/BF01911126
80. Samko S.G. Fractional integration and differentiation of variable order: an overview // Nonlinear Dynamics. 2013. Vol. 71. No. 4. P. 653–662. DOI 10.1007/s11071-012-0485-0
81. Kiryakova V. A brief story about the operators of the generalized fractional calculus // Fractional Calculus and Applied Analysis. 2008. Vol. 11. No. 2. P. 203–220.
82. Kiryakova V.S. Generalized Fractional Calculus and Applications. New York: Longman and J. Wiley, 1994. 360 p.
83. Luchko Yu., Trujillo J.J. Caputo-type modification of the Erdelyi-Kober fractional derivative // Fractional Calculus and Applied Analysis. 2007. Vol. 10. No. 3. P. 249–268.
84. Prudnikov A.P., Brychkov Yu.A., Marichev O.I. Integrals and Series. Vol. 1: Elementary Functions. New York: Gordon and Breach, 1986. 798 p.
85. Kiryakova V., Luchko Yu. Riemann-Liouville and Caputo type multiple Erdelyi-Kober operators // Central European Journal of Physics. 2013. Vol. 11. No. 10. P. 1314–1336. DOI: 10.2478/s11534-013-0217-1
86. Caputo M. Mean fractional-order-derivatives differential equations and filters // Annali dell'Universita di Ferrara. 1995. Vol. 41. No. 1. P. 73–84. DOI: 10.1007/BF02826009
87. Bagley R.L., Torvik P.J. On the existence of the order domain and the solution of distributed order equations - Part I // International Journal of Applied Mathematics 2000. Vol. 2. P. 865–882.

88. Bagley R.L., Torvik P.J. On the existence of the order domain and the solution of distributed order equations - Part II // International Journal of Applied Mathematics 2000. Vol. 2. P. 965–987.
89. Caputo M. Distributed order differential equations modelling dielectric induction and diffusion // Fractional Calculus and Applied Analysis. 2001. Vol. 4. No. 4. P. 421–442.
90. Lorenzo C.F., Hartley T.T. Variable order and distributed order fractional operators // Nonlinear Dynamics. 2002. Vol. 29. No. 1. P. 57–98. DOI: 10.1023/A:1016586905654
91. Jiao Z., Chen Y.Q., Podlubny I. Distributed-Order Dynamic Systems: Stability, Simulation, Applications and Perspectives. London: Springer, 2012. 90 p. DOI: 10.1007/978-1-4471-2852-6
92. Nahushev A.M. Fractional Calculus and its Application. Moscow: Fizmatlit, 2003. 272 p. [in Russian]
93. Pskhu A.V. On the theory of the continual integro-differentiation operator // Differential Equations. 2004. Vol. 40. No. 1. P. 128–136. DOI: 10.1023/B:DIEQ.0000028722.41328.21
94. Pskhu A.V. Partial Differential Equations of Fractional Order. Moscow: Nauka, 2005. 199 p. [in Russian]
95. Li C., Zeng F. Numerical Methods for Fractional Calculus. Boca Raton: Chapman and Hall / CRC Press, 2015. 281 p.
96. Guo B., Pu X., Huang F. Fractional Partial Differential Equations and Their Numerical Solutions. Singapore: World Scientific, 2015. 336 p.
97. Anastassiou G.A., Argyros I.K. Intelligent Numerical Methods: Applications to Fractional Calculus. New York: Springer, 2016. 423 p. DOI: 10.1007/978-3-319-26721-0
98. Harrod R. An Essay in dynamic theory // Economic Journal. 1939. Vol. 49 (193). P. 14-33.
99. Domar E.D. Capital expansion, rate of growth and employment // Econometrica. 1946. Vol. 14. No. 2. P. 137-147.
100. Domar E.D. Expansion and employment // The American Economic Review. 1947. Vol. 37. No. 1. P. 34-55.
101. Gorenflo R., Kilbas A.A., Mainardi F., Rogosin S.V. Mittag-Leffler Functions, Related Topics and Applications. Berlin: Springer-Verlag, 2014. 443 p. DOI: 10.1007/978-3-662-43930-2
102. Tarasova V.V., Tarasov V.E. Dynamic intersectoral models with power-law memory // Communications in Nonlinear Science and Numerical Simulation. 2018. Vol. 54. P. 100-117. DOI: 10.1016/j.cnsns.2017.05.015
103. Tarasov V.E., Tarasova V.V. Time-dependent fractional dynamics with memory in quantum and economic physics // Annals of Physics. 2017. DOI: 10.1016/j.aop.2017.05.017
104. Umarov S. Introduction to Fractional and Pseudo-Differential Equations with Singular Symbols. Heidelberg: Springer, 2015. 434 p. DOI: 10.1007/978-3-319-20771-1
105. Malinowska A.B., Odzijewicz T., Torres D.F.M., Advanced Methods in the Fractional Calculus of Variations. Heidelberg: Springer 2015. 135 p. DOI: 10.1007/978-3-319-14756-7